\documentclass[Physsubmission, Phys]{SciPost}

\hypersetup{
    colorlinks,
    linkcolor={red!50!black},
    citecolor={blue!50!black},
    urlcolor={blue!80!black}
}

\usepackage[bitstream-charter]{mathdesign}
\DeclareSymbolFont{usualmathcal}{OMS}{cmsy}{m}{n}
\DeclareSymbolFontAlphabet{\mathcal}{usualmathcal}

\begin{document}


\begin{center}{\Large \textbf{
Timelike Transition Form Factor for \\  CP-odd Higgs Boson Production
\\
}}\end{center}

\begin{center}
Ken Sasaki\textsuperscript{1},
Tsuneo Uematsu\textsuperscript{2$\star$}
\end{center}

\begin{center}
{\bf 1}  Dept. of Physics, Faculty of Engineering, Yokohama National University,\\ Yokohama 240-8501, Japan
\\
{\bf 2} Graduate School of Science, Kyoto University, Kitashirakawa, Sakyo-ku, 
\\ Kyoto 606-8502, Japan
\\
* uematsu@scphys.kyoto-u.ac.jp

\end{center}

\begin{center}
\today
\end{center}


\definecolor{palegray}{gray}{0.95}
\begin{center}
\colorbox{palegray}{
  \begin{tabular}{rr}
  \begin{minipage}{0.1\textwidth}
    \includegraphics[width=35mm]{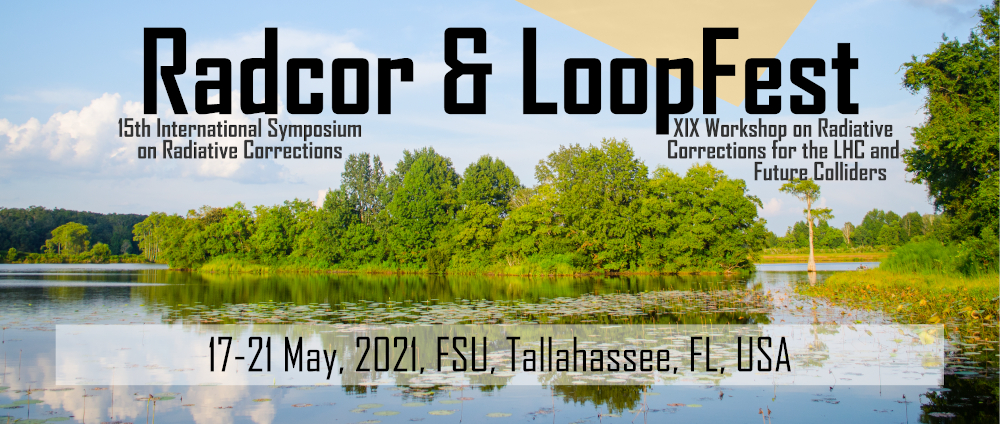}
  \end{minipage}
  &
  \begin{minipage}{0.85\textwidth}
    \begin{center}
    {\it 15th International Symposium on Radiative Corrections: \\Applications of Quantum Field Theory to Phenomenology,}\\
    {\it FSU, Tallahasse, FL, USA, 17-21 May 2021} \\
    \doi{10.21468/SciPostPhysProc.?}\\
    \end{center}
  \end{minipage}
\end{tabular}
}
\end{center}

\section*{Abstract}
{\bf


We investigate the production of CP-odd Higgs boson $A^0$ 
associated with a real $\gamma$ in $e^+e^-$ collisions.
Because of the properties of $A^0$ coupling to the other fields,
the main contribution comes from the top-quark triangle loop 
diagrams.
We obtain the \lq\lq timelike\rq\rq transition form factor 
which describes the production
amplitude of $A^0$ in the $e^+e^-$ collisions. 
This timelike transition form factor is related to the
spacelike transition form factor relevant for the $A^0$
production in  $e^-\gamma$ collisions. It turns out that
the possible extra contributions from chargino-sneutrino
and neutralino-selectron box diagrams do not give sizable effects.
}

\vspace{10pt}
\noindent\rule{\textwidth}{1pt}
\tableofcontents\thispagestyle{fancy}
\noindent\rule{\textwidth}{1pt}
\vspace{10pt}

\section{Introduction}
\label{sec:intro}
The Higgs boson with mass about 125 GeV was discovered by ATLAS and CMS 
at LHC~\cite{HiggsLHC} and its spin, parity and couplings were examined~\cite{SpinParity}. Now it would be intriguing to study its nature in $e^+e^-$ 
collisions provided by linear colliders~\cite{ILC}, which would offer much 
cleaner experimental data. In our previous works, we have studied the production of the SM Higgs boson ($ H_{\rm SM}$)~\cite{KWSUPL,WKUSPRD} and the CP-odd Higgs boson ($A^0$)~\cite{SU,SU2} in $e\gamma$ collisions at one-loop level in the electroweak interaction, where $A^0$ appears in the type-II 2HDM or in the MSSM~\cite{Hunter}. There we have  particularly focused on the transition form factor (TFF) of the production of $H_{\rm SM}$ and $A^0$ bosons which is the analog of the TFF of the $\pi^0$ production in $e^+e^-$ collisions.

In the present talk, we investigate the production of the CP-odd Higgs 
boson $A^0$ associated with a real $\gamma$ in $e^+e^-$ collisions. 
It turns out that the main contribution, at one-loop level, comes from the 
top-quark triangle loop diagrams, because of the properties of $A^0$ coupling 
to the other fields. We obtain the \lq\lq timelike\rq\rq  TFF which describes the production amplitude of $A^0$ in $e^+e^-$ collisions. We show that this timelike TFF is related to the spacelike 
 TFF appearing in the $A^0$ production in the $e\gamma$ 
collisions. We find that the possible extra contributions from 
chargino-sneutrino and neutralino-selectron box diagrams do not 
give sizable effects, within the parameter space we consider.

\section{Spacelike and timelike transition form factors}
\begin{figure}[hbt]
 \begin{center}
 \includegraphics[scale=0.25]{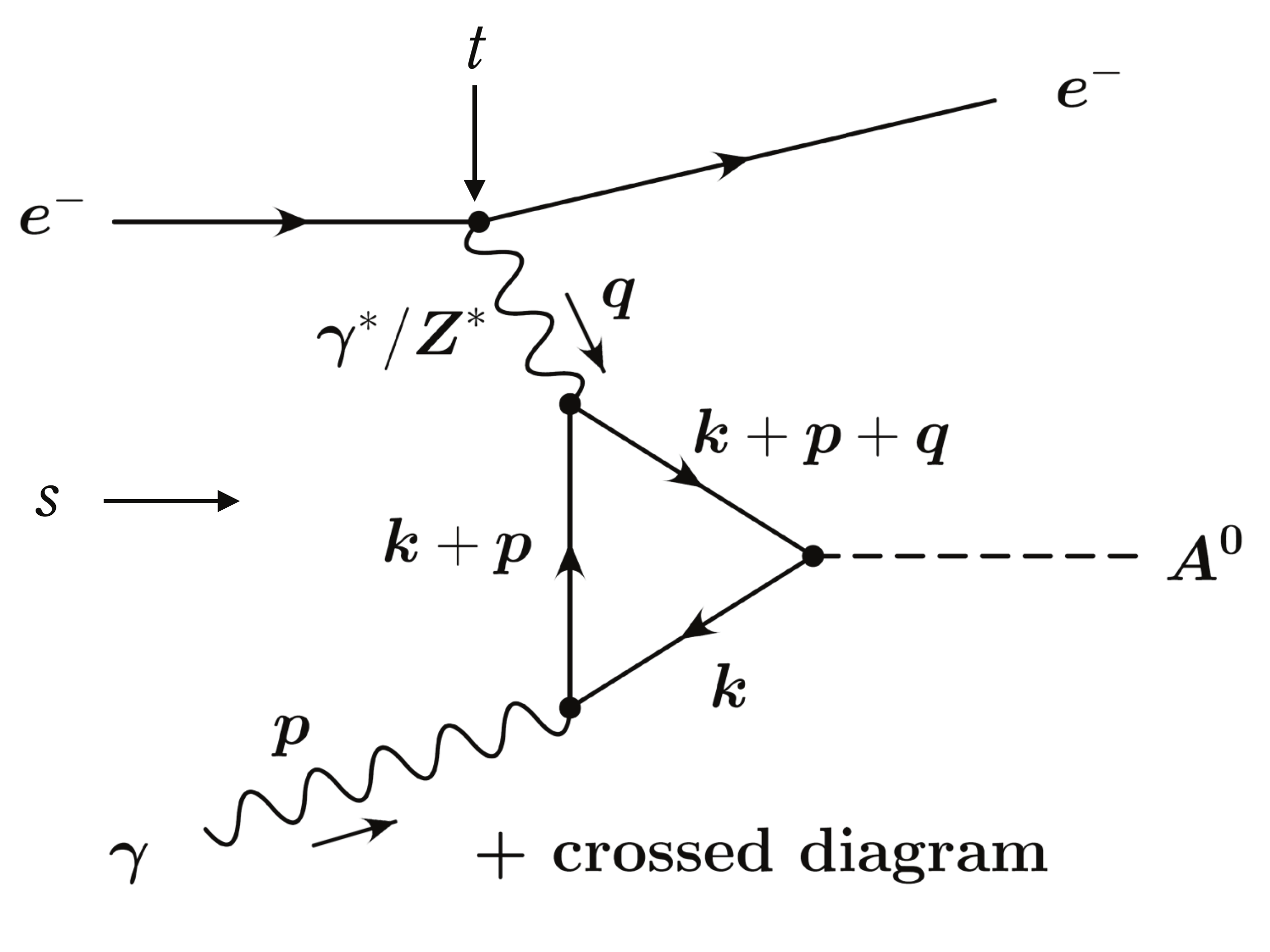}
 \hspace{-0.1cm}
 \includegraphics[scale=0.25]{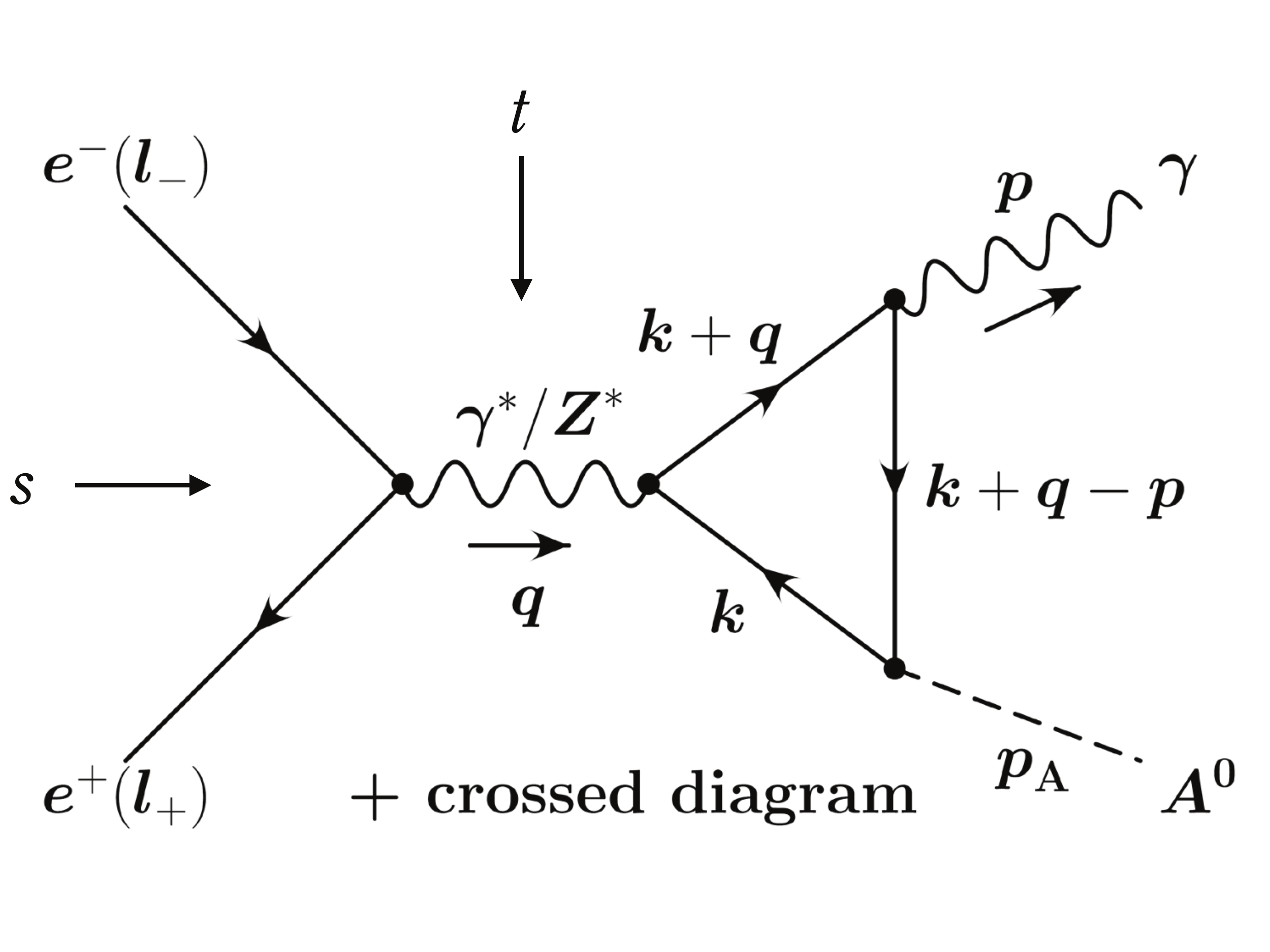}\\
 \vspace{-0.1cm}
 (a) \hspace{5.5cm}(b)
 \end{center}
 \vspace{-0.3cm}
 \caption{\label{egamma-epluseminus} 
 CP-odd Higgs $A^0$ production in (a) $e\gamma$ and  (b) $e^+e^-$
collisions.
 }
\end{figure}

Let us consider two processes of $A^0$ production mediated by virtual 
$\gamma^*/Z^*$ exchanges in $e\gamma$ and $e^+e^-$ collisions, 
which are related to each other 
by exchanging Mandelstam variables $s$ and $t$ as shown in 
Fig.\ref{egamma-epluseminus}. 
The contributions of the top-quark loop diagrams to the amplitudes for these processes are written in terms of $A^t_{\mu\nu}$ which is expressed as follows:
\begin{eqnarray}
A^t_{\mu\nu}=-\frac{e^2g}{(4\pi)^2}N_Cq_t^2\frac{\cot\beta}{2m_W}F_t(\rho,\tau)
\varepsilon_{\mu\nu\rho\sigma}q^\rho p^\sigma~,
\end{eqnarray}
where $e$ and $g$ are the electromagnetic and weak gauge couplings, respectively, and $m_W$ is the $W$ boson mass, $q_t$ is the charge of top-quark, $N_c$ is the number of colors, 
and $\cot\beta=v_1/v_2$ is the ratio of the vacuum expectation values of the two Higgs 
doublets in the type-II 2HDM or in the MSSM. 
The function $F_t(\rho,\tau)$ corresponds to the relevant TFF which arises from the top-quark loop.

In our previous work on $e\gamma\to eA^0$ process~\cite{SU}, we introduced the 
spacelike TFF which is given by
\begin{eqnarray}
F_t^{\ e\gamma\rightarrow eA^0}(\rho,\tau)=\frac{\tau}{1+\rho\tau}
[g^{e\gamma}(\rho)+4f(\tau)]~,
\end{eqnarray} 
where the scaling variables 
$\rho$ and $\tau$ are defined as 
\begin{eqnarray}
\rho\equiv \frac{-q^2}{4m_t^2}, \quad \tau\equiv
\frac{4m_t^2}{m_A^2}\quad (q^2<0)~,
\end{eqnarray}
with $m_t$ ($m_A$) being the top-quark ($A^0$ boson) mass, and the 
function $g^{e\gamma}(\rho)$ is given by
\begin{eqnarray}
g^{e\gamma}(\rho)&=&\left[\log\frac{\sqrt{\rho+1}+\sqrt{\rho}}
{\sqrt{\rho+1}-\sqrt{\rho}}\right]^2~, \label{gspace}
\end{eqnarray}
and $f(\tau)$ is the well-known function which appears in the Higgs decay process $H_{\rm SM}\rightarrow \gamma\gamma$  \cite{Hunter}.

For $e\gamma \to e A^0$ process, $q^2$ is 
negative, while $q^2(=\sqrt{s})$ is positive for the process of $e^+e^-\to A^0\gamma$.  For the latter case 
we  introduce a timelike TFF as follows:
\begin{eqnarray}
F_t^{\ e^+e^-\rightarrow \gamma A^0}(\rho,\tau)=\frac{\tau}{1-\rho\tau}
[-g^{e^+e^-}(\rho)+4f(\tau)]~,\label{TFF-top}
\end{eqnarray}
where $g^{e^+e^-}(\rho)$ turns out to be
\begin{eqnarray}
g^{e^+e^-}(\rho)=-\left[\log\frac{\sqrt{\rho}+\sqrt{\rho-1}}
{\sqrt{\rho}-\sqrt{\rho-1}}-i\pi\right]^2~,\label{gtime}
\end{eqnarray}
and the scaling variables $\rho$ and $\tau$ are now defined as
\begin{eqnarray}
\rho\equiv \frac{q^2}{4m_t^2}, \quad \tau\equiv
\frac{4m_t^2}{m_A^2}\quad (q^2>0)~.
\end{eqnarray}
In the expressions of (\ref{gspace}) and 
(\ref{gtime}), we make an analytic continuation by putting $+i\epsilon$ to 
$q^2$.
\begin{eqnarray}
g^{e\gamma}\left(\frac{-q^2}{4m_t^2}-i\epsilon\right)
=-g^{e^+e^-}\left(\frac{q^2}{4m_t^2}+i\epsilon\right)~,
\end{eqnarray}
which leads to the following relation between the timelike and spacelike TTFs 
given as
\begin{eqnarray}
F_t^{\ e\gamma\rightarrow eA^0}(-\rho,\tau)&=&\frac{\tau}{1-\rho\tau}
[g^{e\gamma}(-\rho)+4f(\tau)]=\frac{\tau}{1-\rho\tau}
[-g^{e^+e^-}(\rho)+4f(\tau)]\nonumber\\
&=&F_t^{\ e^+e^-\rightarrow \gamma A^0}(\rho,\tau)~.
\label{relation}
\end{eqnarray}

\begin{figure}[hbt]
 \begin{center}
\includegraphics[scale=0.28]{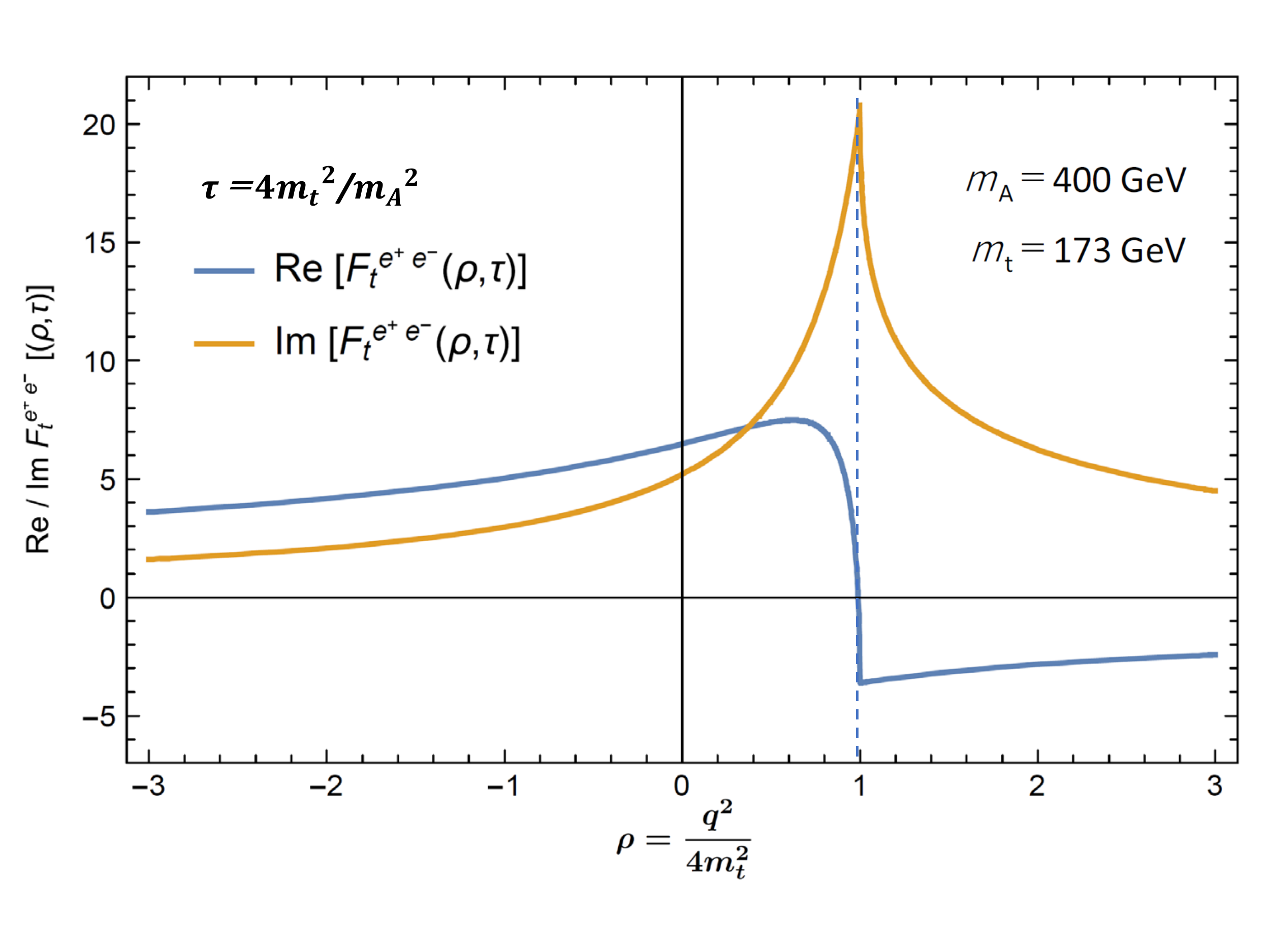}
 \hspace{-0.1cm}
 \includegraphics[scale=0.28]{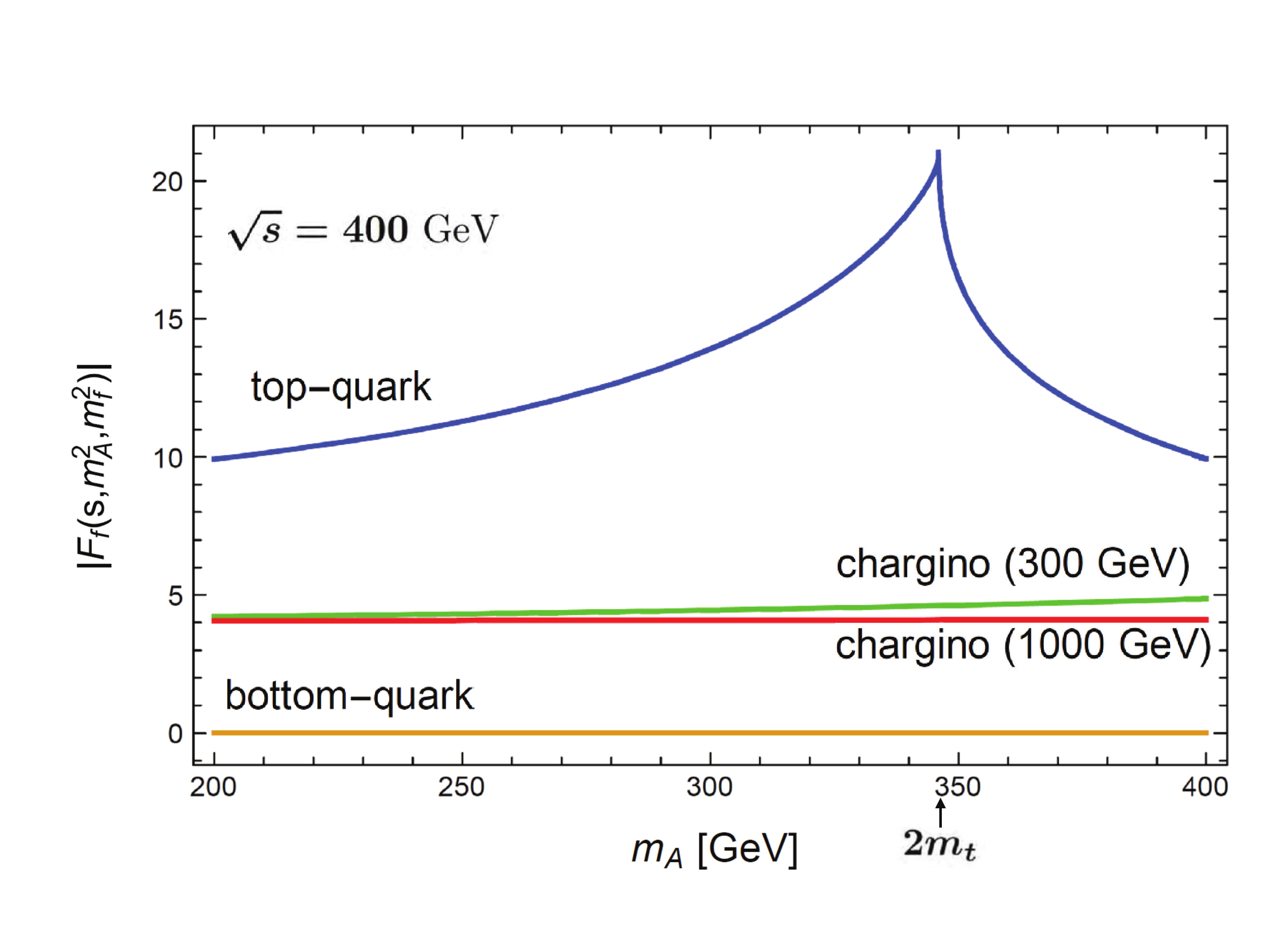}\\
 \vspace{-0.1cm}
 (a) \hspace{6.5cm}(b)
 \end{center}
 \vspace{-0.3cm}
 \caption{\label{TFF-fns} 
Transition form factor for $A^0$ production in
$e^+e^-$ collisions:(a) Real and imaginary parts of $F_t^{e^+e^-}(\rho,\tau)$
, and 
(b) Absolute value of $F_f^{e^+e^-}(\rho,\tau)$ for various fermion
loops; top-quark, bottom-quark, chargino with mass 300 GeV and 1000 GeV.
  }
\end{figure}

In Fig.\ref{TFF-fns}(a) we plot the real and imaginary parts of the TFF in the whole region from spacelike $q^2<0$ to timelike $q^2>0$ as a function of $\rho=q^2/4m_t^2$. As can be seen from the figure, the imaginary part (orange line) exhibits a cusp structure at $\rho=1$, which corresponds to the 
$t\bar{t}$ threshold, {\it i.e}, $q^2=(2m_t)^2$, while the real part (blue line) changes sign at $\rho=1$.
The production cross section depends on the absolute square of the 
 TFF and thus it shows a cusp structure at $\rho=1$ (see Fig.\ref{total-cross}(a) and (b)).
In Fig.\ref{TFF-fns}(b), we show contributions to the 
absolute value of transition form factor $|F_f(s,m_A^2,m_f^2)|$ from 
the triangle one-loop diagram of fermion $f$; 
top-quark, bottom-quark and  chargino with mass 300 GeV and 1000 GeV. As seen from the figure, the chargino and bottom-quark do not give sizable effects. The main contribution 
comes from the top-quark loop diagram. We also note that squarks 
do not contribute, since the trilinear $A^0$ coupling 
to  mass-eigenstate  squark pairs vanishes~\cite{SU}.  

\section{CP-odd Higgs $A^0$ production in $e^+e^-$ collisions}

As a minimal extension of the Higgs sector of the SM, we consider the type-II 2HDM which includes the MSSM as a special case~\cite{Hunter}. After the spontaneous symmetry breaking, there appear two charged $H^\pm$ and three neutral  $h^0$, $H^0$ (CP-even), $A^0$ (CP-odd) Higgs bosons.
The characteristics of $A^0$  couplings to other fields 
 are as follows : 1) In contrast to the CP-even Higgs bosons 
$h^0$ and $H^0$, $A^0$ does not couple to $W^+W^-$ and $ZZ$ pairs 
at tree level. Hence $W$- and $Z$-boson one-loop diagrams do not contribute to the $A^0$ production. 
2) $A^0$ does not couple to other two physical Higgs bosons in  cubic interactions. 3) The couplings of $A^0$ to quarks and leptons are proportional to their masses. Therefore, we only consider the top-quark for the charged fermion loop diagrams. The $A^0$ coupling to the top-quark with mass $m_t$ is given by $\lambda \gamma_5$ with $ \lambda =-gm_t\cot\beta/(2m_W)$.

We investigate the $A^0$ production process: 
$e^-(l_-) +e^+(l_+) \rightarrow \gamma(p) +A^0(p_A)$,
where we detect the produced real photon in the final state.
This process has been studied already in the literature 
\cite{Djouadi-etal}. Here we reconsider the process in the light
of transition form factors.
The one-loop diagrams which contribute to the reaction
are classified into two groups: $\gamma^*$-exchange and $Z^*$-exchange diagrams (see Fig.\ref{egamma-epluseminus}(b)).
As we will see later, the contribution of the 
former is far more dominant over that of the latter.
The scattering amplitude for each process turns out to be
\begin{eqnarray}
&&\hspace{-1.3cm}\langle\gamma A^0|T|e^-e^+\rangle^t_{\gamma^*}=
[\overline{v}(l_+)(-ie\gamma^\mu)u(l_-)]\frac{-i}
{q^2+i\epsilon}A_{\mu\nu}^t\epsilon^\nu(p)~,\label{VirtualGamma}\\
&&\hspace{-1.3cm}\langle \gamma A^0|T|e^-e^+\rangle^t_{Z^*}
=\frac{g}{4\cos\theta_W}[\overline{v}(l_+)
(i\gamma^\mu)(f_{Ze}+\gamma_5)u(l_-)]\frac{-i}{q^2-m_Z^2+i\epsilon}\widetilde{A}^t_{\mu\nu}\epsilon^\nu(p)~,\label{VirtualZ}
\end{eqnarray}
with
\begin{eqnarray}
&&A_{\mu\nu}^t=-\frac{e^2g}{(4\pi)^2}N_Cq_t^2\frac{\cot\beta}{2m_W}F_t(\rho,\tau)
\varepsilon_{\mu\nu\rho\sigma}q^\rho p^\sigma~,
\\
&&\widetilde{A}_{\mu\nu}^t=-\frac{eg^2}{(4\pi)^2}\frac{N_Cq_tf_{Zt}}{4\cos\theta_W}\frac{\cot\beta}{2m_W}F_t(\rho,\tau)\varepsilon_{\mu\nu\rho\sigma}q^\rho p^\sigma~,
\end{eqnarray}
where $u(l_-)$ and $\overline{v}(l_+)$ are the spinors for the electron and 
positron with momentum $l_-$ and $l_+$, respectively,  $\epsilon^\nu(p)$ 
is the produced photon polarization vector with $p_\nu\epsilon^\nu(p)=0$, and 
$f_{Ze}=-1+4\sin^2\theta_W$ and $f_{Zt}=1-(8/3)\sin^2\theta_W$ are the strength of vector part of the $Z$-boson coupling to electron and top-quark, 
respectively, with $\theta_W$ being the Weinberg angle.


Adding two amplitudes given in Eqs.(\ref{VirtualGamma}) and (\ref{VirtualZ}), we calculate  the differential cross section for the $A^0$ production in $e^+e^-$ collisions with unpolarized initial beams, which is expressed by the sum of three terms:
\begin{eqnarray}
&&\hspace{-1.3cm}\left(\frac{d\sigma}{dt}\right)_{\gamma^*}=\frac{\alpha_{\rm em}^3}{64\pi}
\frac{g^2}{4\pi}\Bigl(\frac{\cot\beta}{2m_W}\Bigr)^2\frac{1}{s}\left(\frac{t^2+
u^2}{s^2}\right)
\Bigl|N^t_Cq_t^2F_t(q^2,m_A^2, m_t^2)\Bigr|^2~,\label{gamma-gamma-cross}
\\
&&\nonumber\\
&&\hspace{-1.3cm}\left(\frac{d\sigma}{dt}\right)_{Z^*}=\frac{\alpha_{\rm em}}{64\pi}
\Bigl(\frac{g^2}{4\pi}\Bigr)^3\Bigl(\frac{\cot\beta}{2m_W}\Bigr)^2\Bigl( \frac{1}{16\cos^2\theta_W} \Bigr)^2\frac{s}{(s-m_Z^2)^2}
\nonumber\\
&& \times f_{Zt}^2(f_{Ze}^2+1) \left(\frac{t^2+u^2}{s^2}\right)\Bigl|N^t_Cq_tF_t(q^2,m_A^2, m_t^2)\Bigr|^2~,\label{Z-gamma-cross}\\
&&\hspace{-1.3cm}\left(\frac{d\sigma}{dt}\right)_{\rm int}=-2\times \frac{\alpha^2_{\rm em}}{64\pi}
\Bigl(\frac{g^2}{4\pi}\Bigr)^2\Bigl(\frac{\cot\beta}{2m_W}\Bigr)^2 \frac{1}{16\cos^2\theta_W} \frac{1}{s-m_Z^2}\nonumber\\
&&\times f_{Zt} f_{Ze}q_t \left(\frac{t^2+u^2}{s^2}\right) \Bigl|N^t_Cq_t
F_t(q^2,m_A^2, m_t^2)\Bigr|^2~,\label{interference-cross}
\end{eqnarray}
where each
corresponds to the contribution of the $\gamma^*$-exchange diagrams, the  $Z^*$-exchange diagrams and their interference, 
respectively, and $\alpha_{\rm em}=e^2/4\pi$.

\section{Numerical analysis of the cross sections}
\begin{figure}[hbt]
 \begin{center}
 \includegraphics[scale=0.28]{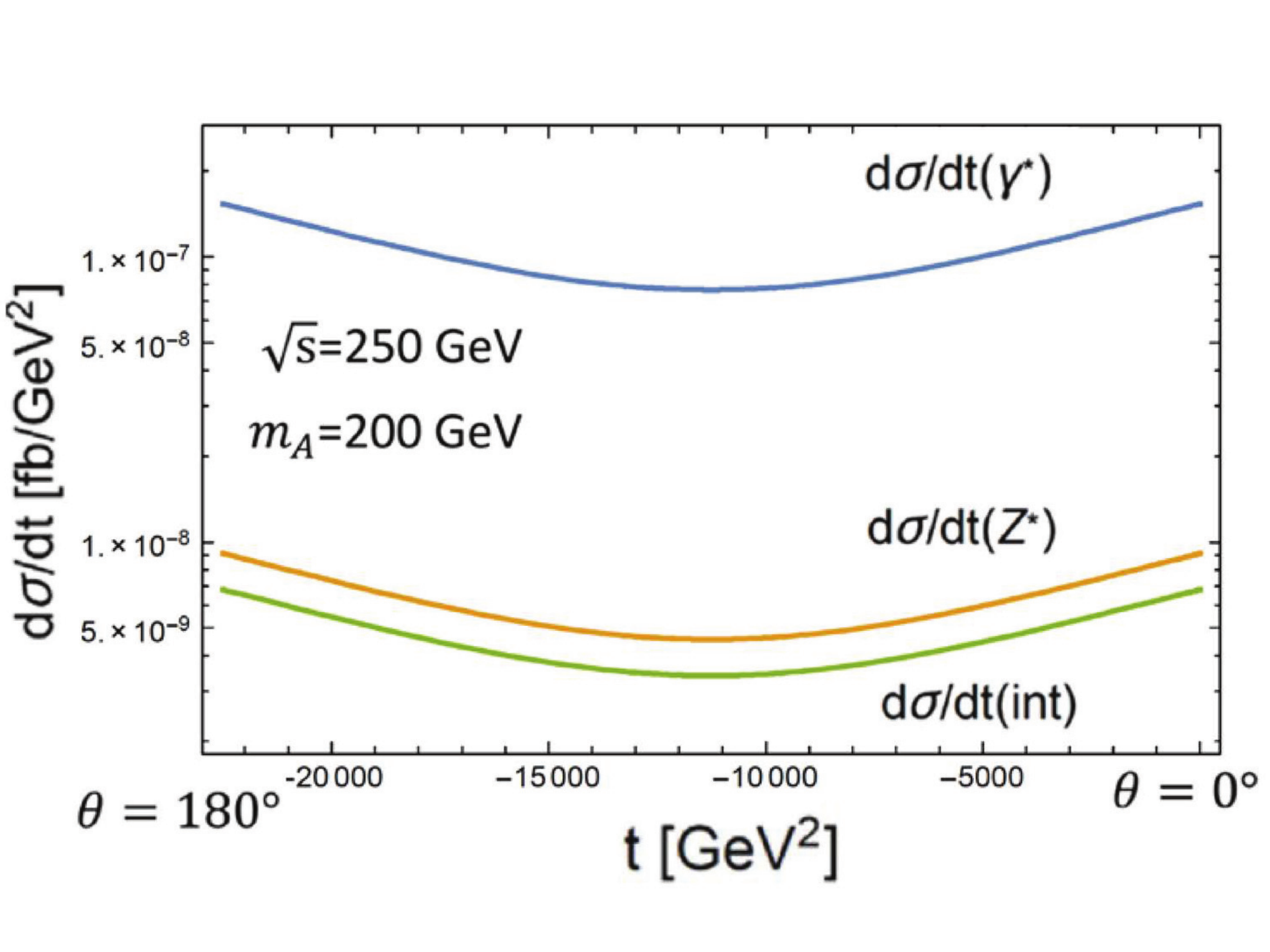}
 \hspace{-0.1cm}
 \includegraphics[scale=0.28]{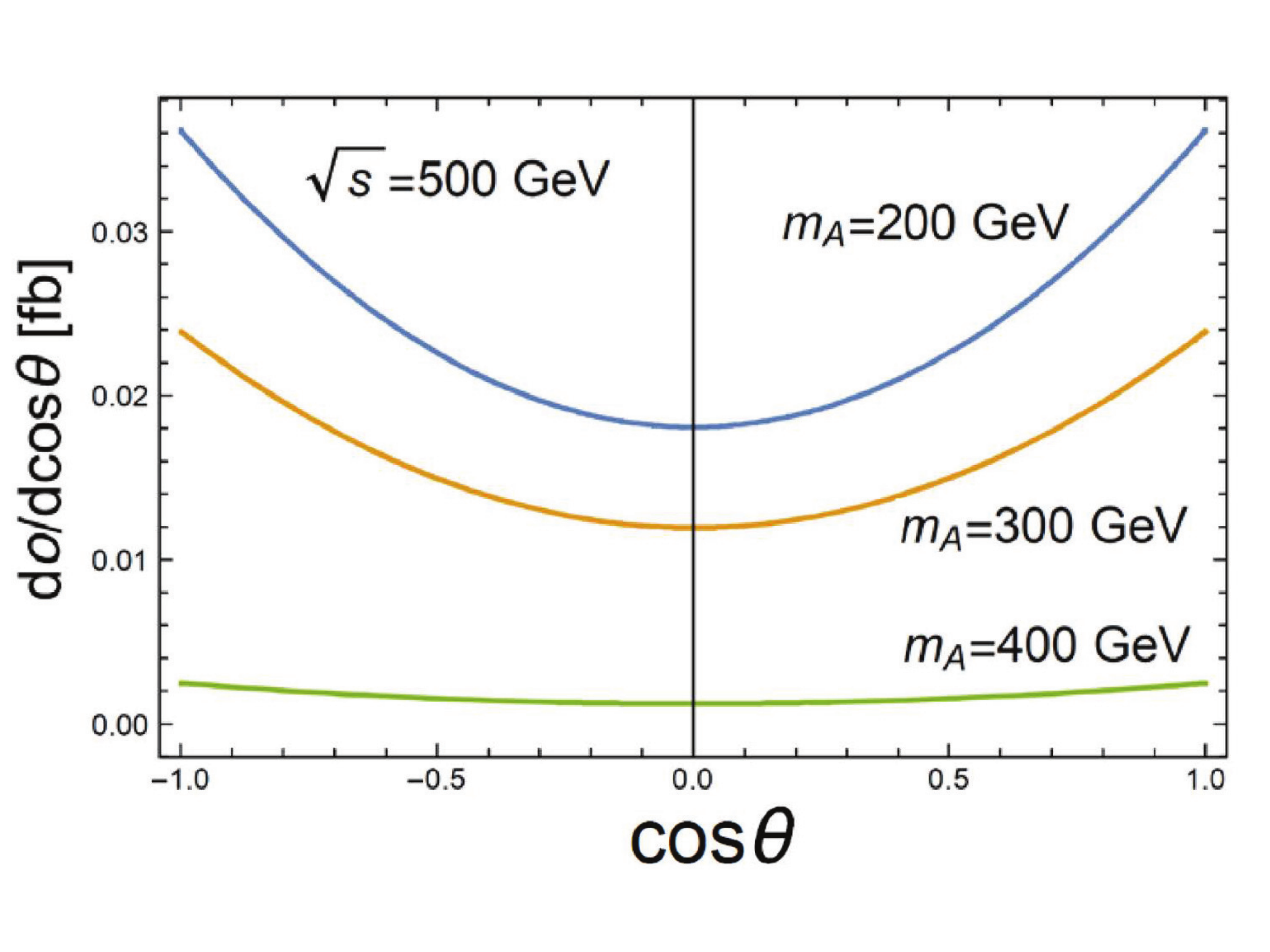}\\
 \vspace{-0.1cm}
 (a) \hspace{6.5cm}(b)
 \end{center}
 \vspace{-0.3cm}
 \caption{\label{diff-cross} 
Differential cross section for $A^0$ production in
$e^+e^-$ collisions:(a) Three components of cross section and 
(b) $A^0$ mass dependence of differential cross section.
  }
  \end{figure}

\begin{figure}[hbt]
 \begin{center}
 \includegraphics[scale=0.28]{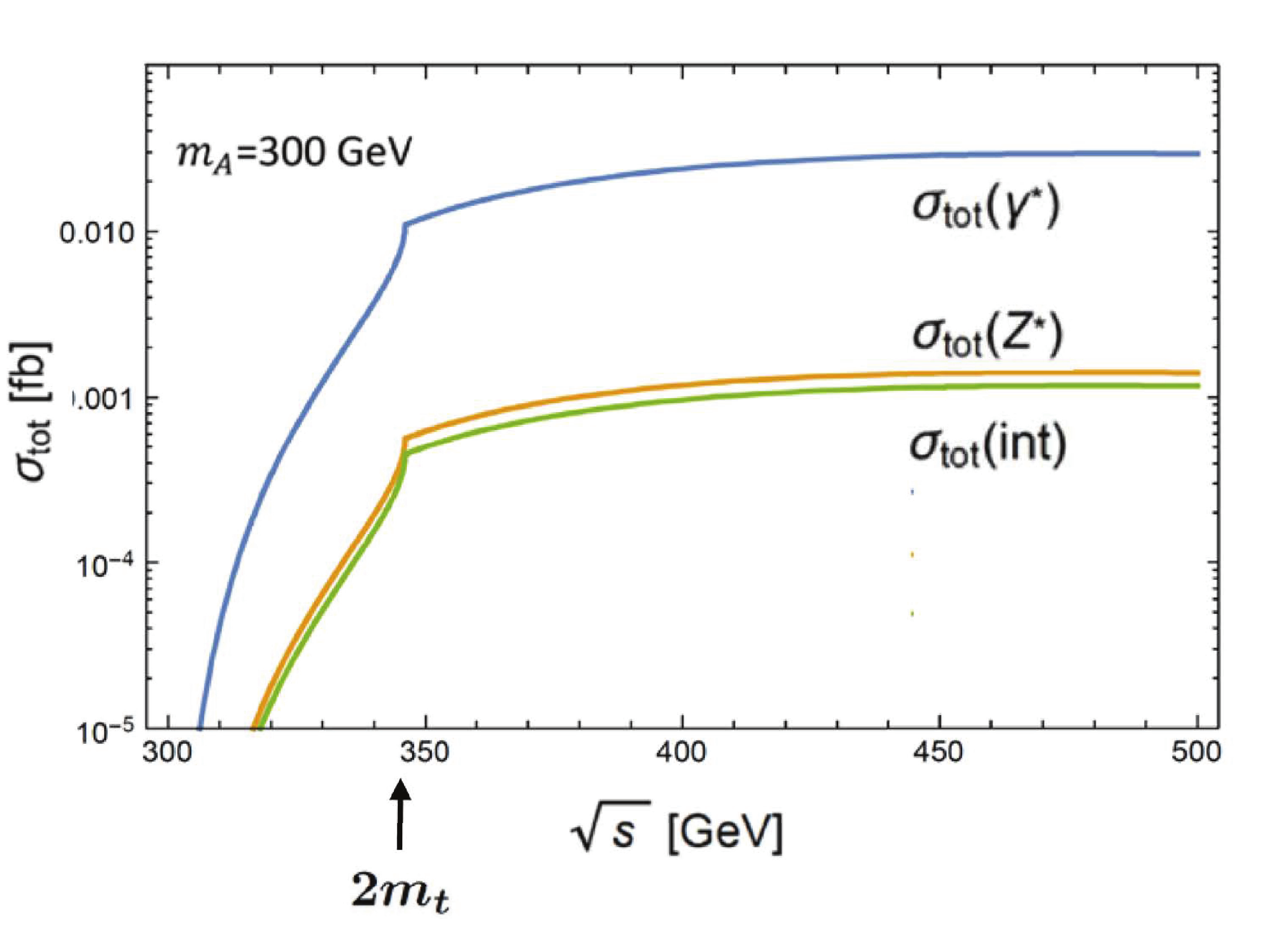}
 \hspace{-0.1cm}
 \includegraphics[scale=0.28]{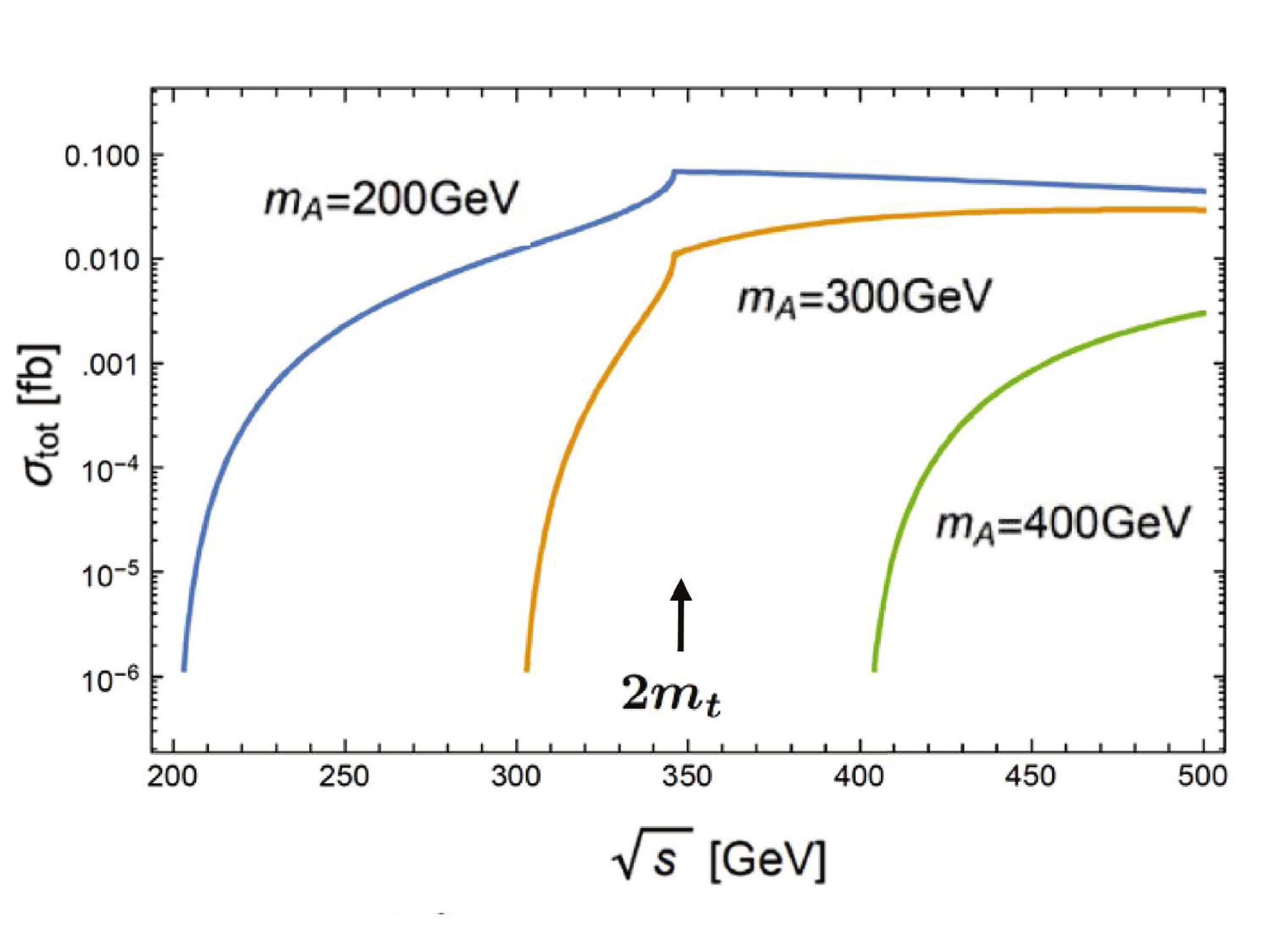}\\
 \vspace{-0.1cm}
 (a) \hspace{6.5cm}(b)
 \end{center}
 \vspace{-0.3cm}
 \caption{\label{total-cross} 
Total cross section for $A^0$ production in
$e^+e^-$ collisions:(a) Three components of cross section and 
(b) $A^0$ mass dependence of total cross section. }
\end{figure}

We analyze numerically the three differential cross sections given in Eqs.(\ref{gamma-gamma-cross})-(\ref{interference-cross}). We choose the parameters
as follows: $m_t=173~{\rm GeV}$, $m_Z=91~{\rm GeV}$, 
$m_W=80~{\rm GeV}$, $\cos\theta_W={m_W}/{m_Z}$, $e^2=4\pi\alpha_{em}=
{4\pi}/{128}$, $g={e}/{\sin\theta_W}$.
The electromagnetic constant $e^2$ is chosen to be the value at the scale of 
$m_Z$. 
For the above choice of parameters, we find $ f_{Zt} f_{Ze}<0$ and, therefore,  Eq.(\ref{interference-cross}) shows that the interference between the $\gamma^*$- and $Z^*$-exchange diagrams work constructively and 
${d\sigma_{({\rm Int})}}/{dt}$ is positive.
For the remaining parameters, we assume $m_A$ between 200 GeV and 500 GeV and the moderate values for $\tan\beta$, which fall within
the allowed region for the $(m_A,\tan\beta)$ parameter space \cite{PDG}.
We plot these differential cross sections as a function of $t$ in 
Fig.\ref{diff-cross}(a) for the case
$\sqrt{s}=250$GeV, $m_A=200$GeV and $\cot\beta=1$. Note that the cross sections are proportional to $\cot^2\beta$.
We find that the contribution from the s-channel $\gamma^*$-exchange diagrams 
is far more dominant over the ones from $Z^*$-exchange diagrams as well as from 
the interference term.
We observe that at $\sqrt{s}=250$ GeV, $m_A=200$ GeV \ 
($\sqrt{s}=500$ GeV, $m_A=300$ GeV), the ratio of
$d\sigma_{Z^*}/dt$ to
$d\sigma_{\gamma^*}/dt$ at $\theta=90^\circ$ is 
$5.94\times 10^{-2}\ (4.98\times 10^{-2})$ 
and $d\sigma_{{\rm int}}/dt$ to
$d\sigma_{\gamma^*}/dt$ at $\theta=90^\circ$
is $4.43\times 10^{-2}$\ $(3.98\times 10^{-2})$. 
As for the total cross section, for $m_A=300$ GeV, 
the ratio $\sigma_{\rm tot}(Z^*)/
\sigma_{\rm tot}(\gamma^*)$ ($\sigma_{\rm tot}({\rm int})/\sigma_{\rm tot}(\gamma^*)$)is $4.97\times 10^{-2}$ ($4.06\times 10^{-2}$) as shown in 
Fig.\ref{total-cross}(a).
We  also show
the $m_A$ dependence of the differential and total cross sections 
in Fig.\ref{diff-cross}(b) and in Fig.\ref{total-cross}(b), respectively.

\section{Box diagram contributions}
\begin{figure}[hbt]
 \begin{center}
 \includegraphics[scale=0.28]{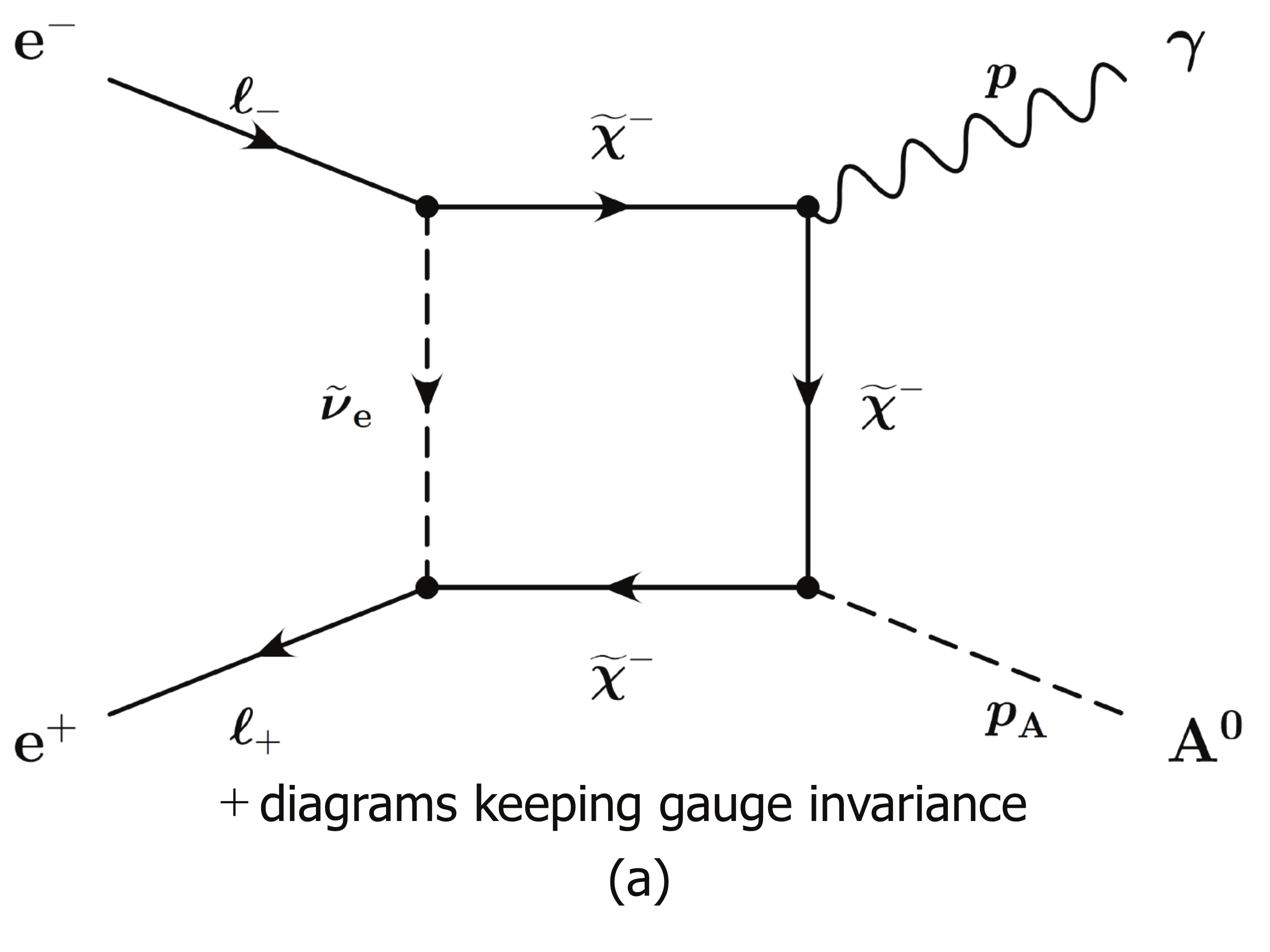}
 \hspace{-0.1cm}
 \includegraphics[scale=0.28]{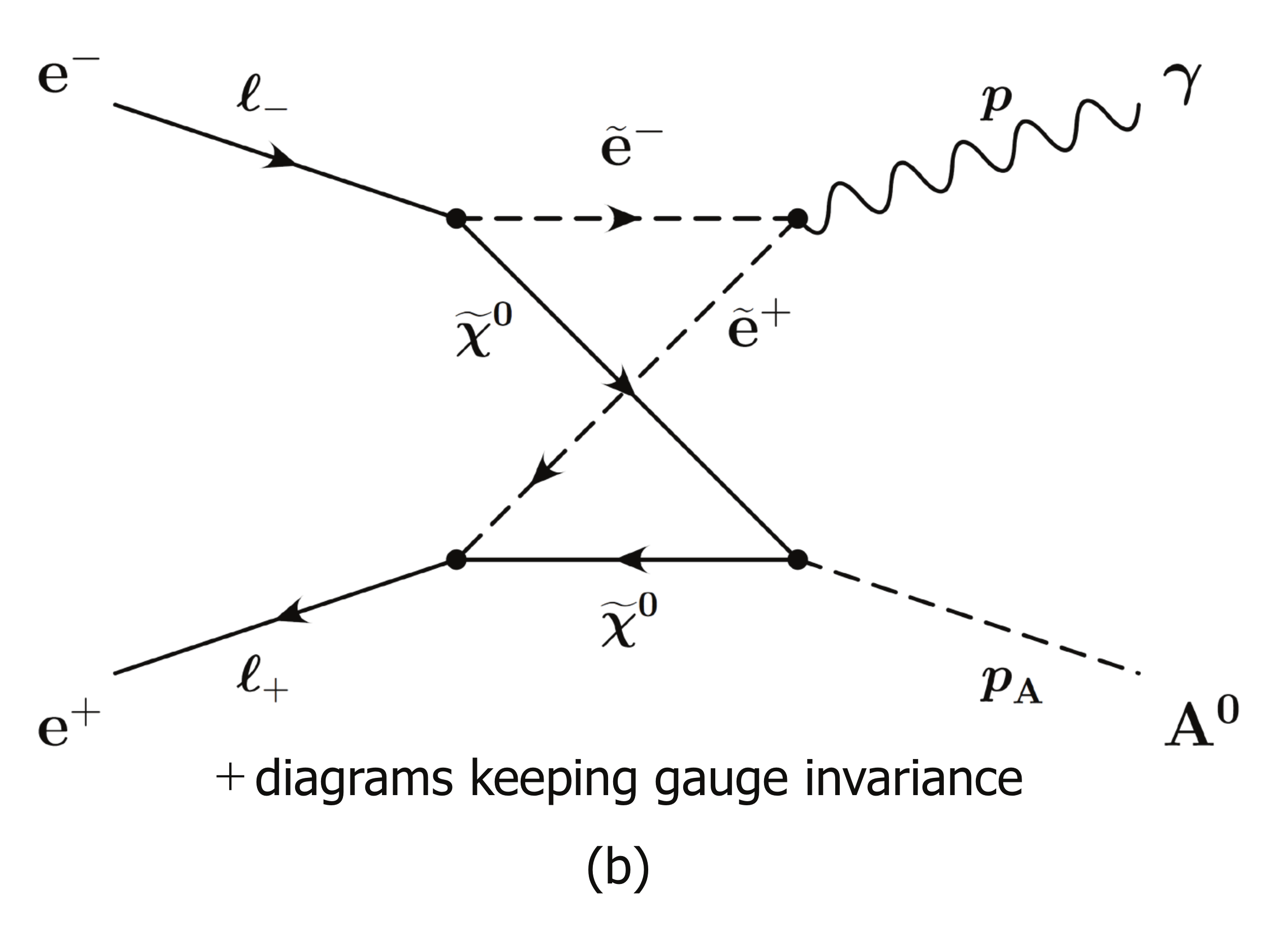}\\
 \vspace{-0.1cm}
 \end{center}
 \vspace{-0.3cm}
 \caption{\label{Box-diagram} Possible box-diagrams in the MSSM:
(a)Chargino($\tilde{\chi}^-$)-sneutrino($\tilde{\nu}_e$)-box contribution and 
(b)Neutralino($\tilde{\chi}^0$)-selectron($\tilde{e}^-$)-box contribution.
}
\end{figure}

Now let us consider under what conditions the TFF
interpretation provides a good framework to describe the process. 
If the contribution from the triangle-loop diagrams is dominant over the one from the box-type diagrams, 
then the TFF leads to a good description of the process, where the production 
cross section is proportional to the absolute square of the TFF.
The possible contributions coming from the box-type diagrams in
 the MSSM are the following: (a) Chargino-sneutrino box diagram
 and (b) Neutralino-selectron 
process shown in Fig.\ref{Box-diagram}. 
The former is the supersymmetric counterpart of the $W$-$\nu_e$ box
diagram and the latter  is the one of $Z$-electron box diagram,  both of which are 
relevant in the case of $H_{\rm SM}\gamma$ production in $e^+e^-$ collisions. 
Of course it is difficult to cover all the parameter spaces for the masses and 
couplings of charginos and sneutrinos and for those of neutralinos and selectrons. 

The one-loop box-diagram amplitude involving chargino $\tilde{\chi}_i^+$ 
($i=1,2$) is given by
\begin{eqnarray}
A_{e^+e^-\rightarrow \gamma A^0}^{(\tilde{\chi}\tilde{\nu})}=
\left(\frac{eg^3\kappa_1|V_{11}|^2}{16\pi^2}\right)\frac{m_{\tilde{\chi}}}{4}
\epsilon^*(p)^{\beta}\left[\overline{v}(l_+)F_{(\tilde{\chi}\tilde{\nu})\beta}(1-\gamma_5)u(l_-)\right]~.
\end{eqnarray}
Since we are only interested in the order of magnitude for the $A^0$ production, we assume that the only mass eigenstate $\tilde{\chi}_1^\pm$
contributes dominantly and take $\kappa_1|V_{11}|^2\sim {\cal O}(1)$  for the coupling of $\tilde{\chi}_1^\pm$ to $A^0$. We can calculate the form factor 
$F_{(\tilde{\chi}\tilde{\nu})\beta}$ in terms of Passarino-Veltman's scalar 
integrals\cite{PassarinoVeltman}. This is obtained by interchanging $s$ and $t$ variables for the case of $e\gamma\to eA^0$ process~\cite{SU2}.
Similar argument holds for the neutralino-selectron box contribution.

\begin{figure}[hbt]
 \begin{center}
 \includegraphics[scale=0.28]{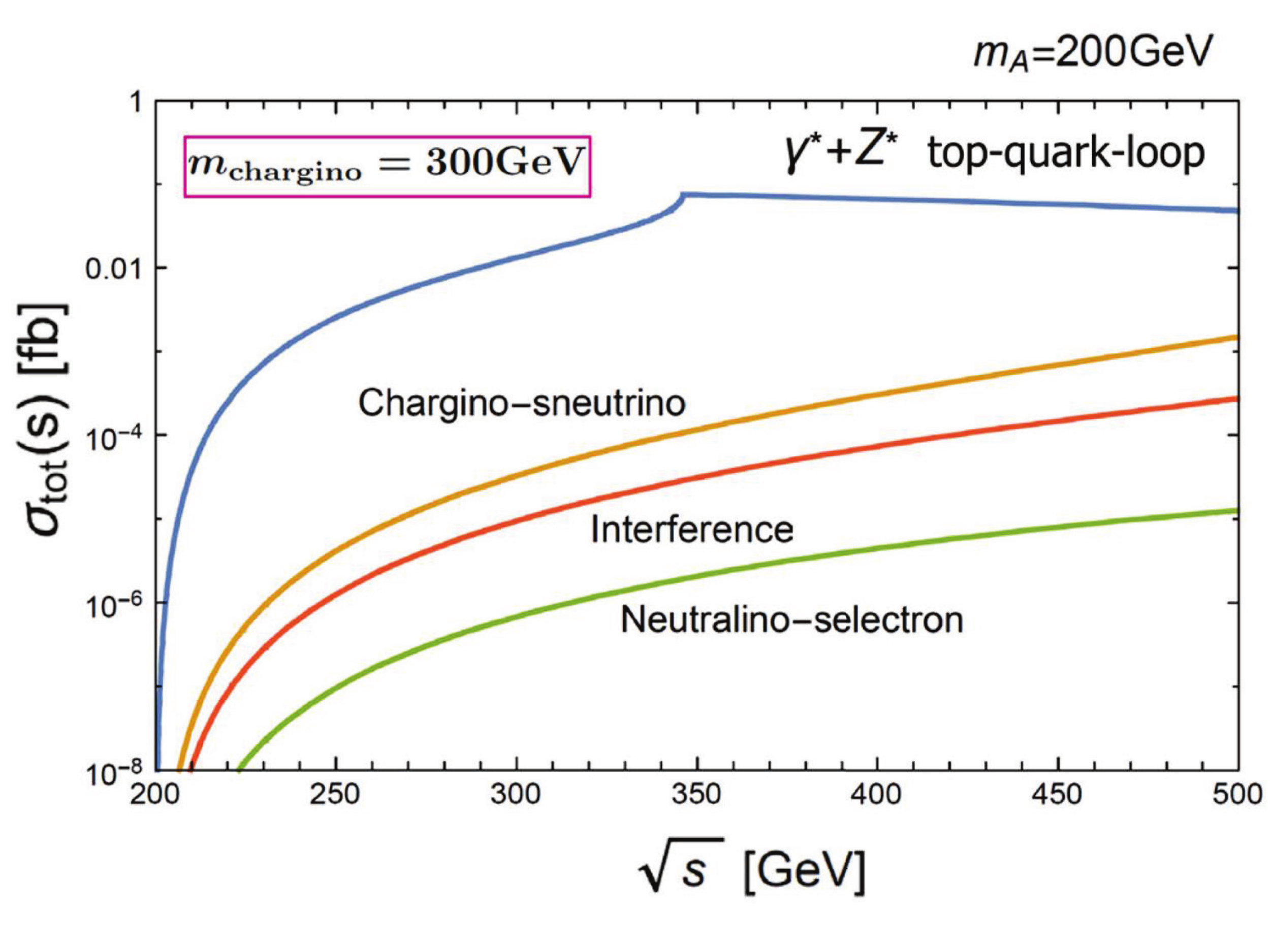}
 \hspace{-0.1cm}
 \includegraphics[scale=0.28]{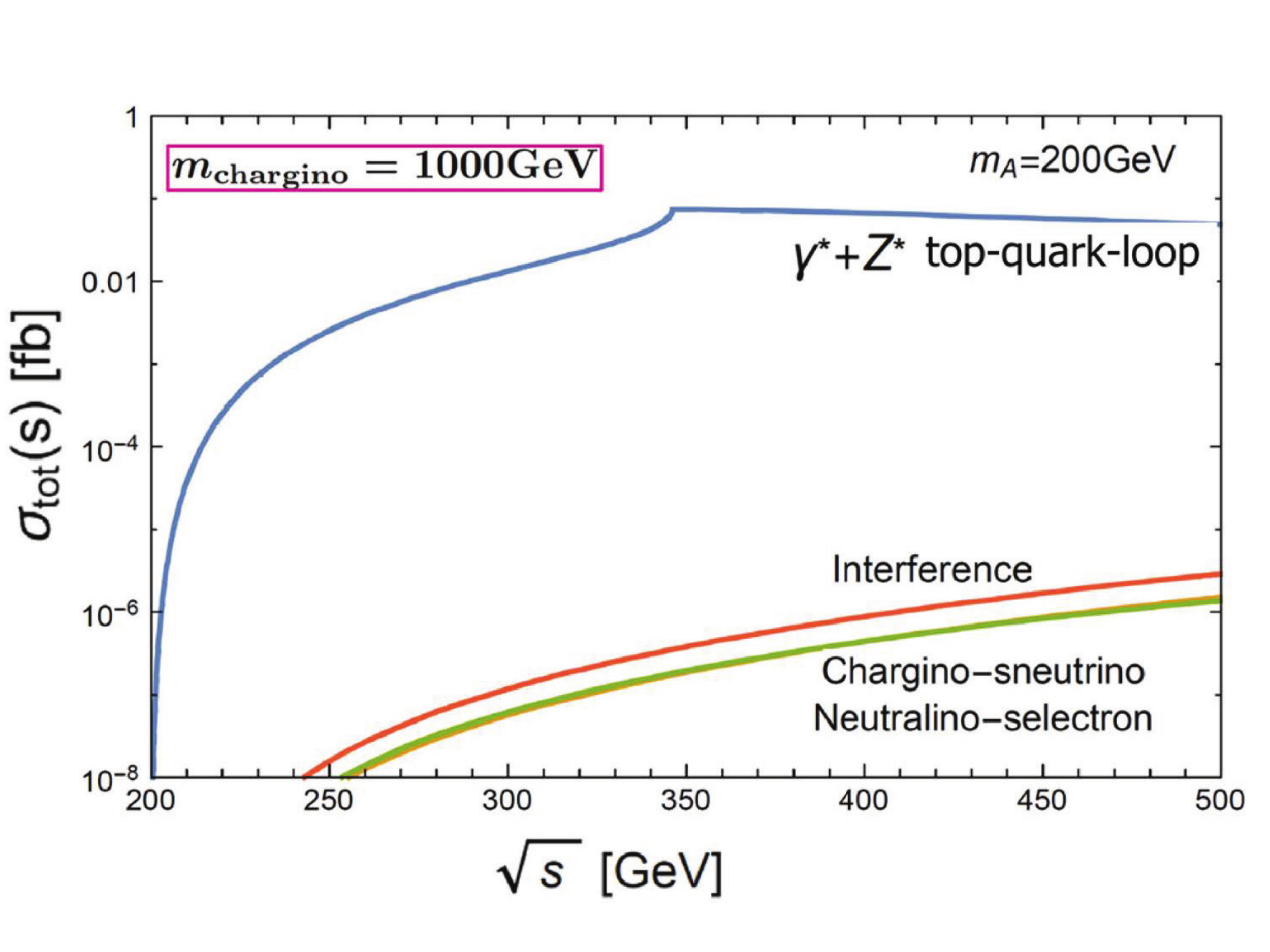}\\
 \vspace{-0.1cm}
 (a) \hspace{6.5cm}(b)
 \end{center}
 \vspace{-0.3cm}
 \caption{\label{Top-vs-Box-diagram} 
Top-quark vs. box-diagrams contributions:(a)
mass of chargino 300 GeV and (b)mass of chargino 1000 GeV.
}
\end{figure}

We show in Fig.\ref{Top-vs-Box-diagram} the total cross section for the 
$A^0$ production in $e^+e^-\to A^0\gamma$ and compare 
the contribution from the box diagrams with the one from the top-quark-loop  
diagrams.
For  chargino with  mass 300 GeV the chargino-sneutrino 
contribution is smaller than the one from top-quark by two order of magnitude. 
With chargino mass 1000 GeV the chargino-sneutrino 
contribution becomes  much smaller by four order of magnitude. In addition 
the interference between top-quark-loop and box diagrams is found to be very small.
The similar box-diagram contributions for the case $e\gamma\to e A^0$ were 
studied in \cite{SU2}, and they were found to be small.
\vspace{-0.8cm}
\section{Conclusion}
In this talk we have studied the production of CP-odd Higgs boson $A^0$ associated 
with a real photon in $e^+e^-$ collisions in terms of the timelike 
 TFF. The dominant contribution is coming from the 
top-quark one-loop diagrams. The $\gamma^*$ process is far more dominant 
over $Z^*$ process. The box-diagram contributions, from chargino-sneutrino 
and neutralino-selectron related processes, do not give sizable effects in 
the parameter space of masses and couplings we have studied. 
If $\tan\beta$ is not large and chargino is very heavy their contributions 
are negligible. Then at the electroweak one-loop level, 
the TFF  provides a good
description of the process. Finally, what about the QCD radiative 
corrections to the top-quark one-loop diagram?  As is well known, for 
$m_t\gg m_A$, 
though it does not apply to our case, the QCD corrections to  the 
effective  $A\gamma\gamma$ coupling are absent.
In the case of $e^+e^-\to H_{\rm SM}\gamma$, the NLO QCD effects are known to be rather small~\cite{Sang-etal}, which 
would suggest that a similar situation occurs in our present case.
Of course, more detailed analyses should be carried out.

\section*{Acknowledgements}
We would like to thank the organizers of the RADCOR 2021 for such a well-organized and stimulating symposium.

\end{document}